\documentclass[runningheads]{llncs}

\usepackage[T1]{fontenc}

\usepackage{graphicx}
\usepackage[firstpage]{draftwatermark}
\usepackage{amssymb}

\usepackage{mathtools}
\usepackage{times}
\usepackage{adjustbox}
\usepackage{xcolor}
\usepackage{tikz}
\usepackage{pgfplots}
\usepackage{booktabs}

\usepackage{multirow}
\pgfplotsset{compat=1.18}
\usepackage{cite}
\usepackage{subfigure}
\usepackage[colorlinks=true]{hyperref}
\usepackage[capitalise]{cleveref}

\usetikzlibrary{decorations.pathreplacing,
                calligraphy,
                matrix,
                positioning}

\DeclareMathOperator{\relu}{ReLU}
\DeclareMathOperator{\recu}{H}
\DeclareMathOperator{\sigmoid}{Sigmoid}

\newcommand{\expo}{\mathrm{e}}
\def\real{\mathbb{R}}

\SetWatermarkText{\hspace*{9cm}\raisebox{12.75cm}{\includegraphics{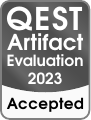}}}
\SetWatermarkAngle{0}

\begin{document}

\title{On the Trade-off Between Efficiency and  Precision of Neural Abstraction}

\author{Alec Edwards\inst{1} 
\orcidID{0000-0001-9174-9962}\and
Mirco Giacobbe\inst{2}
\orcidID{0000-0001-8180-0904}\and
Alessandro Abate\inst{1}
\orcidID{0000-0002-5627-9093}}
\authorrunning{A. Edwards, M. Giacobbe, A.Abate}
\institute{University of Oxford, Oxford, UK \and
University of Birmingham, Birmingham, UK
}
\maketitle

\setcounter{footnote}{0}
\begin{abstract}
Neural abstractions have been recently introduced as formal approximations of complex, nonlinear dynamical models. They comprise a neural ODE and a certified upper bound on the error between the abstract neural network and the concrete dynamical model. So far neural abstractions have exclusively been obtained as neural networks consisting entirely of $\relu$ activation functions, resulting in neural ODE models that have piecewise affine dynamics, and which can be equivalently interpreted as linear hybrid automata. In this work, we observe that the utility of an abstraction depends on its use: some scenarios might require coarse abstractions that are easier to analyse, whereas others might require more complex, refined abstractions. We therefore consider neural abstractions of alternative shapes, namely either piecewise constant or  nonlinear non-polynomial (specifically, obtained via sigmoidal activations). We employ formal inductive synthesis procedures to generate neural abstractions that result in dynamical models with these semantics. Empirically, we demonstrate the trade-off that these different neural abstraction templates have vis-a-vis their precision and synthesis time, as well as the time required for their safety verification (done via reachability computation). We improve existing synthesis techniques to enable abstraction of higher-dimensional models, and additionally discuss the abstraction of complex neural ODEs to improve the efficiency of reachability analysis for these models.
\end{abstract}

\section{Introduction}
\label{sec:intro}

Abstraction is a fundamental process to advance understanding and knowledge. 
Constructing abstractions involves separating important details from those that are unimportant, in order to allow for more manageable analysis or computation of that which was abstracted. This enables the drawing of conclusions that would otherwise be unobtainable (whether analytically or computationally). 
While the features and the properties of abstractions vary between fields and applications, in the context of the analysis and verification of mathematical models of dynamical systems abstractions take on a formal role:  
namely, abstractions must retain all behaviours of the concrete model that they abstract, while at the same time ought to be easier to analyse or do computations on. These requirements allow for certain formal specifications, such as safety, to be formally provable over the abstractions and, additionally, for these very specifications to hold true for the concrete (original) dynamical model (which was the object of abstraction). 

Properties which make an abstraction `good' cannot be universally declared, and instead are deemed to be `useful' in relation to given properties that are the object of analysis.
Abstractions that are `simpler' in structure may be easier to analyse, but might be coarser and thus lead to less refined proofs. In contrast, a more refined abstraction, i.e., one that contains fewer additional behaviours relative to the concrete model, might be complex and challenging to analyse. Indeed, for any given problem (a dynamical model and a property), it is unclear at the outset where the optimal middle ground between simplicity and precision lies, leading to abstraction techniques that begin with coarser ones and refine them iteratively  \cite{clarke2000CounterexampleguidedAbstractionRefinement, tacas/RoohiP016, bogomolov2017CounterexampleGuidedRefinementTemplate}.

The trade-off between this relationship is the topic of this paper: we study how neural abstractions of different nature and precision can be useful for the formal verification of safety properties of complex (e.g., nonlinear) dynamical models. 
Recent literature has introduced the use of neural networks as templates to abstract dynamical models \cite{abate2022neural}. These \emph{neural abstractions} consist of a neural network interpreted as an ODE \cite{nips/ChenRBD18}, alongside a formal (bounded) error between abstract and concrete models. This enables safety verification of models with challenging dynamical behaviours, such as non-linear dynamics including non locally-Lipschitz vector fields. 
While this approach has in principle no constraints on the neural template (e.g., on its shape or on the activation functions employed), in practice they have experimentally been limited to feed-forward neural networks consisting entirely of $\relu$-based activation functions \cite{abate2022neural}. The resulting abstractions are thus piecewise affine models, and can be interpreted as linear hybrid automata~\cite{DBLP:conf/lics/Henzinger96}.
Local linearity is desirable, as the analysis of (piecewise) linear models is a mature area, yet one needs only look at the annual ARCH competition to notice the recent advances in the formal verification of alternative models, such as nonlinear ones, which offer advantages related to their generality. It thus makes sense in this work to explore neural abstractions in a broader context, across different layers of simplicity and precision, as per our discussion above. 

Piecewise constant and nonlinear (Lipschitz continuous) models are well studied model semantics with mature tools for their analyses, each with different advantages. 
In this work, we construct neural-based dynamical models which follow these semantics and cast them as formal abstractions, and show their potential use cases.
While piecewise constant and piecewise affine are clearly simpler templates to abstract nonlinear ODEs, it is perhaps initially unclear how the latter can be considered useful. However, not all nonlinear models are alike in complexity: models with complex functional composition or with the presence of functions that are not Lipschitz continuous will be abstracted to models which are still nonlinear, but free of these constraints. The potential of nonlinear neural abstractions is further bolstered by recent advances in developing literature on reachability analysis for neural ODEs \cite{manzanaslopez2022ReachabilityAnalysisGeneral, aaai/GruenbacherHLCS21, gruenbacher2022}, which are leveraged in this work.

\clearpage

We summarise our main contributions in the following points: 
\begin{itemize}
    \item we build upon a recent approach that leverages neural networks as abstractions of dynamical models \cite{abate2022neural}, by introducing new neural piecewise constant and nonlinear templates (\cref{sec:prelim}), 
    \item we extend and improve existing synthesis procedures to generate abstractions and cast them as equivalent, analysable models (\cref{sec:synth}, \cref{sec:safe_ver}),
    \item we implement the overall procedure\footnote{The codebase is available at \href{https://github.com/aleccedwards/qest-na}{https://github.com/aleccedwards/qest-na.}} under different abstract models, and empirically demonstrate tradeoffs of different templates with regards to precision and speed (\cref{sec:results}),
    \item we also demonstrate the ability of our approach to abstract higher-dimensional models, as well as complex neural ODEs, which is significant in light of the growing field of research around neural ODEs (\cref{sec:results}). 
\end{itemize}

\subsection*{Related Work}
We observe that the verification of dynamical models with neural network structure is an active area of research. In particular, there is much interest in verifying models with continuous-time dynamics and neural network controllers that take actions at discrete time steps, often referred to as neural network control systems \cite{cav/TranYLMNXBJ20,tecs/HuangFLC019,hybrid/DuttaCS19,cav/IvanovCWAPL21,schilling2022}. Whilst in this work we focus on continuous-time models, we emphasise that verification of discrete-time models and of models with neural controllers is also studied \cite{ijcai/BacciG021, tecs/TranCLMJK19, amcc/XiangTRJ18}. 

As previously mentioned, safety verification of nonlinear continuous-time dynamical models is a mature field. Analysis of such models can be performed using Taylor models \cite{rtss/ChenAS12, cav/ChenAS13,  tecs/ChenMS17, rtss/0002S16}, abstractions   \cite{hybrid/Althoff13, cav/MoverCGIT21, hybrid/AsarinD04, hybrid/Sankaranarayanan11, cav/SankaranarayananT11}, simulations \cite{cav/FanQM0D16}, reduction to first-order logic \cite{tacas/KongGCC15}, and via Koopman linearisation  \cite{Bak_Bogomolov_Duggirala_Gerlach_Potomkin_2021}. 
In this work we focus on the challenging setup of verification of nonlinear models with possibly non-Lipschitz vector fields, which violate the working assumptions of all existing verification tools.

In this work we observe that neural ODEs with piecewise constant or affine dynamics induce a partitioning of the state space, which casts them as hybrid automata. The approximation of a (nonlinear) vector field as a hybrid automaton is is know as \emph{hybridisation}, and has been extensively utilised for the analysis of nonlinear models \cite{hybrid/AsarinDG03, hybrid/DangMT10, hybrid/DangT11,  hybrid/BakBHJP16, hybrid/HenzingerW95, hsb/KongBBGHJS16, formats/BogomolovGHK17, cdc/KekatosFF17, formats/LiBB20, rtss/SotoP20, 871304},  even with dynamics as simple as piecewise constant ones \cite{acta/AsarinDG07, cav/PrabhakarS13}. However, typically these approaches rely on a fixed partitioning of the state space (often based on a grid) that is chosen a-priori, which can limit their scalability.  
Hybridisation schemes commonly rely on error-based over-approximation to ensure soundness, though this idea is not exclusive to methodologies that partition the state space \cite{cdc/AlthoffSB08}, and has been formalised by means of the notions of (bi)simulation relations  \cite{pola2008ApproximatelyBisimilarSymbolic, cav/MajumdarZ12, tac/PrabhakarD015}.

\section{Preliminaries}\label{sec:prelim}
In this section we introduce the necessary preliminary material for this work, beginning with familiar concepts of dynamical models and neural networks before moving on to the recently introduced neural abstractions.

\subsection{Dynamical Models and Safety Verification}
This work studies continuous-time autonomous dynamical models which consist of coupled ODEs affected by an uncertain disturbance \cite{sastry1999NonlinearSystems, khalil2002NonlinearSystems}. Denote a dynamical model as 
\begin{equation}\label{eq:dynamical-system} 
    \mathcal{F}: \quad 
    \dot{x} = f(x) + d, 
    \quad \|d\| \leq \delta,
    \quad x \in \mathcal{X}, 
\end{equation} 
where $f: \real^n \to \real^n$ is a nonlinear vector field, and the state vector $x$ lies within some Euclidean domain of interest $\mathcal{X} \subset \real^n$. More formally, we denote the dynamical model in Eq. \eqref{eq:dynamical-system} as  $\mathcal{F}$. 
The disturbance $d$ is a signal with range $\delta > 0$,  
where $\|\cdot\|$ denotes a norm operator 
(unless explicitly stated, we assume all norms 
are given the same semantics across the paper). 
The symbol $d$ is interpreted as a non-deterministic disturbance, which at any time can take any possible value within the disturbance bound provided by $\delta$. 

We study the following formal verification question:  whether a continuous-time, autonomous dynamical model with uncertain disturbances is safe within a region of interest and a given time horizon, and with respect to a region of initial states $\mathcal{X}_0 \subset \mathcal{X}$ and a region of bad states $\mathcal{X}_B \subset \mathcal{X}$.  
Verifying safety consists of determining that no trajectory initialised in $\mathcal{X}_0$ enters $\mathcal{X}_B$ over the given time horizon.

We tackle safety verification by means of formal abstractions, 
followed by reachability analysis via flowpipe (i.e., reach sets) propagation. We construct an abstract dynamical system that captures all behaviours (trajectories, executions) of the original system. We then calculate (sound over-approximations of) the flowpipes of these abstractions: i.e., we find (sound over-approximations of) the reachable states from the initial set $\mathcal{X}_0$. If these states do not intersect with the bad states, we can conclude the abstraction is safe. Since we deal with over-approximation, if the abstract model is safe, then the concrete model is necessarily safe too \cite{hybrid/AsarinDG03}. Since the final check involving set intersection is quite straightforward, the crux of the safety verification problem is the computation of first, the abstract model, and second, the reachable states via flow-pipe propagation - both steps will be the focus of our experimental benchmarks.   
\subsection{Linear Hybrid Automata}
\label{sec:lha}
Hybrid automata are useful models of systems comprising both continuous and discrete dynamics~\cite{DBLP:conf/lics/Henzinger96,van_der_Schaft_Schumacher_2000}. They consist of a collection of (discrete) modes, each with their own dynamics over continuous states, rules for transitioning amongst discrete modes, and consequently to re-initialise the continuous dynamics into a new mode. Different semantics exists for hybrid automata; in this work we use the same semantics as SpaceEx~\cite{alur1995AlgorithmicAnalysisHybrid,cav/FrehseGDCRLRGDM11}, which is a state-of-the-art tool used for analysis of linear hybrid automata.

The present work considers linear hybrid automata induced by a state space partitioning. 
Define a hybrid automaton $\mathcal{H} = (Mod, Var, Lab, Inv, Flow, Trans)$, consisting of a labeled graph encoding the evolution of a finite set of continuous variables $Var \subset \real^n$, in which the state variables $x$ take their values. 
Each vertex $m_i \in Mod$ of the graph is referred to as a mode, and corresponds to a partition within the state space.
The state of the automaton is thus given by the pair $(m_i, x)$, whose evolution in time is described by $Flow(m_i,x)$. 
We describe this evolution as the flow, and in each mode the local flow given by $\dot{x}(t) = f_{m_i}(x) + d$, where $f_{m_i}(x) \in \real^n$ is affine in $x$ and $d$ is a non-deterministic disturbance such that $||d|| \leq \delta$. The edges of the graph describing $\mathcal{H}$ are known as transitions. 
A transition $(m, \alpha, Guard, m') \in Trans$ has label $\alpha \in Lab$ and allows the system state to jump from $(m, x)$ to the successor state $(m', x)$ instantaneously.
We note that only the mode changes during a transition, with $x$ being mapped via the identity function. 
Transitions are governed by $Guard \subset \real^n$, with a transition being enabled when $x \in Guard$.
Moreover, the state may only remain in a mode $m$ if the continuous state is within the invariant $Inv(m) \subset \real^n$. For the purposes of hybrid automata induced by state-space partitioning in this work, both $Guard$ and $Inv(m)$ are closed convex polyhedra $\mathcal{P} = \{x | \bigwedge_{i} a_i \cdot x \leq b_i \}$, where $a_i \in \real^n$ and $b_i \in \real$.
We denote by $\mathcal{P}_{i}$ the polyhedron that defines the invariant for mode $m_i$.

\subsection{Neural Networks}
Let a feed-forward neural network $\mathcal{N}(x)$ consist of an $n$-dimensional input layer $y_0$, 
$k$ hidden layers $y_1, \dots, y_k$ with dimensions $h_1, \dots, h_k$ respectively, and an $n$-dimensional output layer $y_{k+1}$ \cite{mackay2003InformationTheoryInference}.
Each hidden or output layer with index $i$ is respectively associated to matrices of weights $W_i \in \mathbb{R}^{h_i \times h_{i-1}}$, a vector of biases $b_i \in \mathbb{R}^{h_i}$ and a nonlinear activation function $\sigma_i$. 
The value of every hidden layer and the final output $y_{k+1}$ is given by the following equations:
\begin{equation}
\label{eq:layer}
    y_i = \sigma_i(W_i y_{i-1} + b_i), \quad y_{k+1} = W_{k+1}y_k + b_{k+1}. 
\end{equation} 
We remark that the expression used for the activation functions clearly determines the features of a neural network: later we shall discuss three alternatives, and how these affect abstract models built on these ``neural templates''.

\subsection{Neural Abstractions}
The concept of a neural abstraction has been recently introduced in \cite{abate2022neural} and is recalled here as follows.

Consider a feed-forward neural network $\mathcal{N} \colon \mathbb{R}^n \to \mathbb{R}^n$ together with a user-specified error bound $\epsilon > 0$.  
Together, these define a neural abstraction $\mathcal{A}$ for the dynamical model $\mathcal{F}$, which is given by nonlinear function $f$ and 
disturbance radius $\delta$, over a region of interest $\mathcal{X}$, if it holds true that
\begin{equation}
    \label{eq:na-spec}
    \forall x \in \mathcal{X}
    \colon 
    \|f(x) - \mathcal{N}(x)\| \leq \epsilon - \delta.
\end{equation}  
Then, the neural abstraction $\mathcal{A}$ consists of the following dynamical system,  
defined by a neural ODE with bounded disturbance:
\begin{equation}
    \dot{x} = \mathcal{N}(x) + d, 
    \quad \|d\|\leq \epsilon,
    \quad x \in \mathcal{X}.\label{eq:na}
\end{equation}

Altogether, we define the neural abstraction $\mathcal{A}$ of a non-linear dynamical system $\mathcal{F}$ as a neural ODE with an additive 
disturbance, and note that it approximates the dynamics of $\mathcal{F}$, while also accounting for the approximation error. 
Notably, no assumption is placed on the vector field $f$: in particular, $f$ is not required to be Lipschitz continuous. This is because the precision of a neural abstraction relies on the condition described by \eqref{eq:na-spec}, whose certification is performed by an SMT solver (see \cref{sec:synth}). 

The choice of activation function $\sigma$ here is significant, as it determines the structure of the final abstraction, which in turn impacts its quality and utility. 
The work in \cite{abate2022neural} focuses on neural abstractions consisting of a neural network 
with $\relu$ activations, and presents an abstraction workflow and experimental results that depend on this specific choice.  
The $\relu$ activation is a suitable choice in neural nets.  Such functions originate a polyhedral partitioning of the input domain, since each $\relu$ function can be `on' or `off', corresponding to two half-spaces -- one of which has a linear output and the other zero output, 
\begin{equation}
    \label{eq:relu}
    \relu(x) = \begin{cases}
    x, &\text{if } x > 0 \\
    0, &\text{otherwise}.
    \end{cases}
\end{equation}
Any fixed configuration of $\relu$ functions as `on' or `off' can be therefore be interpreted as the intersection of a finite number of halfspaces, which itself is a convex polyhedron.
Furthermore, the output layer of the neural net induces functions that are piecewise affine. 
In light of this, \cite{abate2022neural} shows that a feed-forward neural ODE with $\relu$ activations induces a linear hybrid automaton. Furthermore, it shows how this linear hybrid automaton can be cast from the $\relu$ network, by constructing the labelled graph for $\mathcal{H}$. This involves calculating the appropriate modes (vertices) --- which requires the computation of the convex polyhedral invariants and locally affine flows --- and transitions (edges) --- whose guards are equivalent to the invariant of the mode being transitioned to. Due to space limitations, details of this cast are not discussed and readers are directed to \cite{abate2022neural} instead.

In this work, we consider additional instances of neural abstractions which induce a special kind of affine hybrid automaton, in which the local flows are `constant', as well as `nonlinear' neural abstractions that do not induce any  piecewise state partitioning but just `regularise' the original dynamics.

\subsection{Activation Functions Determine the Shape of the Abstract Neural Models}

Approximations of nonlinear vector fields do not need to be piecewise affine.  
In the present work we consider two additional templates for neural abstractions, resulting in abstract dynamical models of different shapes, namely piecewise constant and nonlinear. We expect the former alternative to be simpler to generate than the piecewise affine one, but arguably less precise (at least if the spatial partitioning is the same). Conversely, nonlinear neural abstractions will be seen as ``regularised'' alternatives of the original nonlinear vector fields. This aligns with our intuition about the trade-off between simplicity and precision, as elaborated in the introduction of this work. We require a specific activation function for each of the two classes, as discussed next.

\paragraph{Piecewise Constant Templates}
Neural piecewise constant templates can be constructed using the unit-step function, also known as the Heaviside function, namely 

\begin{equation}
    \label{eq:recu}
    \recu(x) = \begin{cases}
    1, &\text{if } x > 0 \\
    0, &\text{otherwise}.
    \end{cases}
\end{equation}
It is only necessary for the final hidden layer to use piecewise constant activation functions to obtain a piecewise constant output, with all other hidden layers using a $\relu$ activation function. In other words, to obtain a piecewise constant template we set $\sigma_i(x) = \relu(x), \ i=1,\ldots,k-1$ and $\sigma_{k} = \recu(x)$. A neural network with piecewise constant output, rather than a piecewise affine output, will also be equivalent to a specific linear hybrid automaton, namely one with constant flows in each mode (rather than affine flows).

\paragraph{Nonlinear Templates}
For neural nonlinear templates we utilise the $\sigmoid$ function -- resulting in `sigmoidal' nonlinear templates. $\sigmoid$ serves as a smooth approximation to a step function, 
\begin{equation}
    \label{eq:sig}
    \sigmoid(x) = \frac{1}{1 + \expo^{-x}}.
\end{equation}

\section{Formal Synthesis of Template-Dependent Neural Abstractions}\label{sec:synth}
\begin{figure}[htb]
    \centering
    \resizebox{\textwidth}{!}{
    \input{Figures/overview-matrix}
    }
    \caption{Overview of neural abstraction templates presented in this work, alongside details of their synthesis procedure, interpretation,  and corresponding state-of-the-art verification technologies. Here, LHA I refers to \emph{linear hybrid automata} with polyhedral guards and invariants and constant flows, whereas LHA II refers to the same object but now with affine flows.}
    \label{fig:template-overview}
\end{figure}

In this section we discuss a framework for constructing neural abstractions of different expressivities using inductive training procedures coupled with symbolic reasoning. We then describe how these abstractions can be cast from neural ODEs coupled with an error bound to equivalent models (with identical behaviour) that are well studied, alongside mature tools for the analysis of each object. This information is summarised in Fig. \ref{fig:template-overview}.

We leverage an iterative synthesis procedure known as counterexample-guided inductive synthesis (CEGIS, \cite{solar-lezama2006CombinatorialSketchingFinite}), which has been shown to be useful for inductively constructing functions that satisfy desired specifications. It consists of two phases: a training phase which seeks to generate a candidate using data, and a sound certification phase which seeks to (dis)prove this candidate using symbolic reasoning. We now consider each phase in turn.

\subsection{Training Procedures for Neural Abstractions}\label{sec:learning}
\paragraph{Gradient-based Training}
Training neural abstractions for the piecewise affine and the nonlinear templates can be done using gradient descent algorithms (for which we use PyTorch, \cite{pytorch}), which depend on the loss functions used. We seek to  minimise the maximum error over the domain; however, this would result in a non-differentiable loss function, which is undesirable for gradient descent. Therefore, we train via a proxy: the mean squared error,
\begin{equation}\label{eq:loss}
    \mathcal{L}(s) = \frac{1}{|S|} \sum_{s \in S}\|f(s) - \mathcal{N}(s)\|_2.
\end{equation} 
Here $\|\cdot\|_2$ represents the $l^2\text{-norm}$ of its input, and $S$ represents a finite set of data points $s$ that are sampled over $\mathcal{X}$.

\paragraph{Gradient-free Training}
The step-function $\recu(x)$ (cf. Eq. \eqref{eq:recu}) has zero gradient almost everywhere: this makes training with gradient descent-based approaches unsuitable.  
We therefore rely on gradient-free methods for training neural networks, which notably scale less well and converge less quickly. Several options exist as suitable choices: here we use a particle swarm optimisation approach, specifically with PySwarms \cite{pyswarmsJOSS2018}.

Gradient-based training approaches benefit from differentiable loss functions as they ensure gradient calculations are always possible.
However, in the case of neural abstraction synthesis this means we do not train based our true objective, but via a proxy. 
With a non-gradient based approach there are no longer advantages to a differentiable loss function, meaning we can minimise a loss function that better represents our desired specification (cf. Eq. \eqref{eq:na-spec}) based on the maximum error over samples. In line with this, we also utilise the $l^\infty$-norm, namely   
\begin{equation}
\label{eq:loss-max}
    \mathcal{L}'(s) = \max_{s \in S}\|f(s) - \mathcal{N}(s)\|_{\infty}.  %
\end{equation} 

\subsection{Certification of the Quality of a Neural Abstraction}\label{sec:errver}
Regardless of the template used to generate the abstraction, the procedure for the certification of its quality is identical. It involves using an SMT solver to check that at no point in the entire domain of interest $\mathcal{X}$ is the maximum error between the neural network and the concrete model greater than some given $\epsilon$. This upper bound $\epsilon$ on the abstraction error is first estimated empirically using the finite data set $S$. Next, the SMT solver is asked to find an assignment $cex$ of $x$, such that the following formula is satisfiable: 
\begin{equation}
    \label{eq:smt-spec}
    \phi = \exists x \in \mathcal{X} \colon \ \|f(x) - \mathcal{N}(x)\| > \epsilon - \delta.
\end{equation} 

If any such assignment is found then the error bound $\epsilon$ does not hold over the entire domain $\mathcal{X}$, and a valid abstraction has not been constructed. Instead, the assignment is treated as a ``counterexample'' and is added to the data set $S$, as further training of the network continues.  
On the other hand, if no assignment is found, then the specification in Eq. \eqref{eq:na-spec} is valid and the synthesis is complete, returning a sound  neural abstraction. Many suitable choices exist for the SMT solver: we use Z3 \cite{demoura2008Z3EfficientSMT} and DReal \cite{gao2013DRealSMTSolver}. We note that feed-forward neural networks with $\sigmoid$ activation functions are simply elementary functions, and networks with $\relu$ or step-function activations can be interpreted as linear combinations of formulae in first-order logic. These insights enable the encoding of such networks in SMT-formulae by means of symbolic evaluation.

\subsection{Refining the Precision of the Abstraction}\label{sec:error-refine}
The abstraction error given by \cref{eq:smt-spec} is global, that is, it holds across the entire domain $\mathcal{X}$. In reality, the true abstraction error is likely smaller than $\epsilon$ throughout much of the domain. 

As previously stated, piecewise constant and -affine neural networks induce a  hybrid automaton over $\mathcal{X}$; this automaton is realised as part of the proposed framework, after a valid abstraction has been synthesised. Since this automaton induces a polyhedral partitioning of the state-space, we can consider these polyhedra in turn to determine a local abstraction error for each mode. Consider a given mode $m_i$, with flow $f_{m_i}$ and polyhedral invariant $\mathcal{P}_i$. As described in \cref{sec:lha}, the local flow $f_{m_i}$ describes the evolution of $x$ while it remains within the invariant given by $\mathcal{P}_i$.  We can therefore rewrite \cref{eq:smt-spec} as
\begin{equation}
    \label{eq:smt-spec-mode}
    \phi_i = \exists x \in \mathcal{P}_i \colon \ \|f(x) - f(x)_{m_i}\| > \epsilon_i - \delta,
\end{equation}  
where $\epsilon_i \leq \epsilon$ is a candidate upper bound on the abstraction error estimated empirically over the set $S_i \coloneqq \{s \in S | s \in \mathcal{P}_i \}$. 

We provide $\phi_i$ to an SMT solver as before: if no satisfying assignment is found then the candidate error bound holds for the mode. Note that this does not effect the overall correctness of our abstractions: if a satisfying assignment is found, the original upper bound on abstraction error still holds. We present results on the efficacy of this procedure in~\cref{sec:app-b}.

\section{Safety Verification using Neural Abstractions}\label{sec:safe_ver}

Performing safety verification using a neural abstractions amounts to verifying a neural ODE with additive bounded disturbance, as in Equation \eqref{eq:na}.  
A positive outcome of safety verification on the abstract model can, in view of \eqref{eq:na-spec}, 
be directly claimed to also hold for the concrete model in \eqref{eq:dynamical-system}. 
The bottleneck of this step of the safety verification problem is the computation of the reachable states via flow-pipe propagation, which is in general hard for non-linear dynamics but is intended to be mitigated by the `simpler' abstract model: this will indeed be the focus of our experimental benchmarks.   

While recent literature has made advancements in performing reachability analysis on neural ODEs \cite{aaai/GruenbacherHLCS21,gruenbacher2022, manzanaslopez2022ReachabilityAnalysisGeneral}, to the best of our knowledge no such method exists for neural ODEs with an additive bounded disturbance -- a \emph{neural abstraction}. Therefore, rather than handling the neural network directly, we must first construct equivalent models that are amenable to computation. By equivalent, we mean that the trajectories generated by both models are identical. We summarise this information in the final three rows of Fig. \ref{fig:template-overview}.

For the piecewise affine neural abstractions, this involves constructing a hybrid automaton with affine dynamics and invariants defined by polyhedra. The reader is referred to \cite{abate2022neural} for details on this cast, and how this construction can be performed efficiently. The obtained models can be analysed more efficiently using the space-time clustering approximation (STC) algorithm \cite{frehse2013FlowpipeApproximationClustering}, which is implemented in the verification tool SpaceEx \cite{cav/FrehseGDCRLRGDM11}.

We now turn to the newly-introduced piecewise constant models. Noting the similarity between  $\recu(x)$ and $\relu(x)$, neural abstractions with piecewise constant activation functions also induce a linear hybrid automaton, except that now the flows are given by a constant term. Verification of such models is the specialty of the tool PHAVer \cite{frehse2008PHAVerAlgorithmicVerification}, which itself implements a bespoke version of the symbolic model checking algorithm introduced by \cite{alur1996AutomaticSymbolicVerification} and \cite{henzinger1997HYTECHModelChecker}.

Finally, we consider nonlinear neural abstractions with $\sigmoid$ activation functions. We observe that the output of these networks can simply be interpreted as `regularised' nonlinear dynamical models consisting of elementary functions, hence these abstractions are nonlinear ODEs with bounded additive non-determinism. Casting a neural ODE as a nonlinear ODE involves evaluating the network symbolically.
Reachability analysis of these kinds of models can be performed using the mature tool Flow* \cite{cav/ChenAS13}, which performs flowpipe (i.e., reach sets) propagation of nonlinear models using Taylor approximations \cite{rtss/ChenAS12}, and hence is dependent on local Lipschitz continuity. This means that by constructing a neural abstraction with nonlinear templates of a concrete model that is not locally Lipschitz continuous, we enable safety verification (via Flow*) of that otherwise intractable concrete model.

\section{Experimental Results} %
\label{sec:results}
The experiments presented in this section investigate the trade-off between efficiency and precision of neural abstractions.  
We consider a number of benchmarks and study how the expressivity of neural abstractions, which depends on their activation functions, varies across their already-demonstrated niche, namely models that are not locally Lipschitz continuous. We provide the first examination of an additional for which neural abstraction is suitable: that of neural ODEs. 
We demonstrate the efficacy of our abstraction/refinement scheme (cf. \cref{sec:error-refine}), which enables us to consider higher dimensional benchmarks, in~\cref{sec:app-b}.

\subsection{Non-Lipschitz Models}

\begin{table}[htb]
    \centering    
    \caption{Table showing the properties of different abstractions with different templates over a series of benchmarks, 
    and the total time spent for synthesis and flowpipe propagation. Here, `PWC' denotes piecewise constant, `PWA' denotes piecewise affine, 'Sig.' denotes nonlinear sigmoidal; $W$: network architecture (nr. of neurons per layer); $\epsilon$: error bound; $||.||$ denotes the 1-norm, $M$: number of modes in abstraction; $T$: total computation time; $\mu$ denotes average over 10 repeated runs.}

\begin{tabular}{@{} lll rrr rrr rrr @{}}
    \toprule
                            &                   &                  & \multicolumn{3}{c}{$||\epsilon||_1$} & \multicolumn{3}{c}{$M$}   & \multicolumn{3}{c}{$T$}                                                                                              \\ \cmidrule(lr){4-6} \cmidrule(lr){7-9} \cmidrule(lr){10-12}
                            &                   &                  & $\min$                               & \multicolumn{1}{c}{$\mu$} & $\max$                  & $\min$ & \multicolumn{1}{c}{$\mu$} & $\max$ & $\min$ & \multicolumn{1}{c}{$\mu$} & $\max$  \\
    \textbf{Benchmark}      & \textbf{Template} & \textbf{$W$}     &                                      &                           &                         &        &                           &        &        &                           &         \\
    \midrule
    \textbf{Water Tank}     & \textbf{PWC}      & \textbf{[15]}    & 0.16                                 & 0.22                      & 0.27                    & 4      & 5.30                      & 6      & 6.32   & 7.32                      & 7.91    \\
                            & \textbf{PWA}      & \textbf{[12]}    & 0.08                                 & 0.09                      & 0.10                    & 6      & 7.00                      & 8      & 14.55  & 70.37                     & 390.90  \\
                            & \textbf{Sig.}     & \textbf{[4]}     & 0.07                                 & 0.07                      & 0.07                    & 1      & 1.00                      & 1      & 15.53  & 18.31                     & 21.53   \\
    \midrule
    \textbf{Non-Lipschitz1} & \textbf{PWC}      & \textbf{[20]}    & 0.97                                 & 1.21                      & 1.47                    & 14     & 26.50                     & 45     & 13.84  & 16.07                     & 20.56   \\
                            & \textbf{PWA}      & \textbf{[10]}    & 0.10                                 & 0.12                      & 0.14                    & 6      & 14.00                     & 20     & 24.58  & 51.94                     & 92.40   \\
                            & \textbf{Sig.}     & \textbf{[4]}     & 0.05                                 & 0.08                      & 0.10                    & 1      & 1.00                      & 1      & 76.24  & 97.20                     & 105.60  \\
    \midrule
    \textbf{Non-Lipschitz2} & \textbf{PWC}      & \textbf{[24]}    & 1.58                                 & 1.97                      & 2.25                    & 55     & 73.40                     & 94     & 25.81  & 32.97                     & 41.99   \\
                            & \textbf{PWA}      & \textbf{[12 10]} & 0.11                                 & 0.12                      & 0.13                    & 11     & 23.80                     & 62     & 55.94  & 86.47                     & 163.16  \\
                            & \textbf{Sig.}     & \textbf{[10]}    & 0.06                                 & 0.08                      & 0.10                    & 1      & 1.00                      & 1      & 309.82 & 368.36                    & 451.04  \\
    \midrule
    \textbf{Water Tank 4D}  & \textbf{PWC}      & \textbf{[25]}    & 2.47                                 & 2.57                      & 2.69                    & 1      & 1.50                      & 4      & 173.09 & 232.33                    & 400.26  \\
                            & \textbf{PWA}      & \textbf{[12]}    & 0.80                                 & 0.80                      & 0.80                    & 4      & 8.10                      & 18     & 9.14   & 21.49                     & 50.70   \\
                            & \textbf{Sig.}     & \textbf{[7]}     & 0.35                                 & 0.35                      & 0.35                    & 1      & 1.00                      & 1      & 100.08 & 133.00                    & 317.80  \\
    \midrule
    \textbf{Water Tank 6D}  & \textbf{PWA}      & \textbf{[16]}    & 1.30                                 & 1.30                      & 1.30                    & 11     & 16.00                     & 25     & 76.86  & 426.16                    & 1659.32 \\
    \midrule
    \textbf{NODE1}          & \textbf{PWC}      & \textbf{[20]}    & 1.59                                 & 1.83                      & 2.12                    & 33     & 63.10                     & 86     & 36.23  & 70.03                     & 104.95  \\
                            & \textbf{PWA}      & \textbf{[5]}     & 0.20                                 & 0.20                      & 0.20                    & 4      & 5.80                      & 7      & 17.20  & 17.92                     & 18.65   \\
                            & \textbf{Sig.}     & \textbf{[3]}     & 0.20                                 & 0.20                      & 0.20                    & 1      & 1.00                      & 1      & 120.38 & 126.51                    & 137.63  \\
    \bottomrule
\end{tabular}

    \label{tab:results-tables}
\end{table}     

\begin{table}[htb]
    \centering
    \caption{Breakdown of the total computation time shown in \cref{tab:results-tables}. Here, learning time $T_L$; $T_C$: certification time; $T_f$: time to compute flowpipe (i.e., reach sets) over-approximation over horizon of 1 sec, using corresponding verification tool (cf. Fig. \ref{fig:template-overview}).}

\begin{tabular}{@{} lll rrr rrr rrr @{}}
    \toprule
                            &                   & \multicolumn{3}{c}{$T_L$} & \multicolumn{3}{c}{$T_C$} & \multicolumn{3}{c}{$T_f$}                                                                                              \\ \cmidrule(lr){3-5} \cmidrule(lr){6-8} \cmidrule(lr){9-11}
                            &                   & $\min$                    & \multicolumn{1}{c}{$\mu$} & $\max$                    & $\min$ & \multicolumn{1}{c}{$\mu$} & $\max$  & $\min$ & \multicolumn{1}{c}{$\mu$} & $\max$ \\
    \textbf{Benchmark}      & \textbf{Template} &                           &                           &                           &        &                           &         &        &                           &        \\
    \midrule
    \textbf{Water-tank}     & \textbf{PWC}      & 6.16                      & 7.20                      & 7.78                      & 0.02   & 0.03                      & 0.03    & 0.01   & 0.01                      & 0.02   \\
                            & \textbf{PWA}      & 14.43                     & 70.18                     & 390.64                    & 0.02   & 0.05                      & 0.08    & 0.05   & 0.08                      & 0.12   \\
                            & \textbf{Sig.}      & 10.99                     & 13.90                     & 16.79                     & 0.00   & 0.01                      & 0.02    & 3.90   & 4.40                      & 5.13   \\
    \midrule
    \textbf{Non-Lipschitz1} & \textbf{PWC}      & 12.94                     & 14.42                     & 16.91                     & 0.19   & 0.44                      & 0.95    & 0.16   & 0.74                      & 3.22   \\
                            & \textbf{PWA}      & 22.65                     & 43.64                     & 79.17                     & 0.56   & 7.21                      & 20.23   & 0.10   & 0.99                      & 4.97   \\
                            & \textbf{Sig.}      & 19.47                     & 26.63                     & 36.88                     & 0.61   & 1.60                      & 5.17    & 46.01  & 68.97                     & 83.91  \\
    \midrule
    \textbf{Non-Lipschitz2} & \textbf{PWC}      & 11.25                     & 13.75                     & 15.22                     & 1.27   & 1.86                      & 2.58    & 8.51   & 15.97                     & 22.27  \\
                            & \textbf{PWA}      & 28.48                     & 54.20                     & 136.79                    & 12.93  & 23.03                     & 38.76   & 0.45   & 4.41                      & 8.49   \\
                            & \textbf{Sig.}      & 30.73                     & 65.35                     & 127.13                    & 12.76  & 37.17                     & 58.26   & 257.82 & 265.84                    & 275.55 \\
    \midrule
    \textbf{Water-tank-4d}  & \textbf{PWC}      & 172.06                    & 215.69                    & 306.95                    & 0.38   & 0.58                      & 0.94    & 0.20   & 15.84                     & 92.13  \\
                            & \textbf{PWA}      & 6.37                      & 12.63                     & 21.94                     & 2.07   & 8.61                      & 40.64   & 0.09   & 0.13                      & 0.17   \\
                            & \textbf{Sig.}      & 42.67                     & 46.00                     & 51.54                     & 25.98  & 57.09                     & 245.52  & 29.56  & 29.91                     & 30.92  \\
    \midrule
    \textbf{Water-tank-6d}  & \textbf{PWA}      & 38.92                     & 101.84                    & 360.93                    & 16.64  & 323.45                    & 1619.26 & 0.18   & 0.26                      & 0.34   \\
    \midrule
    \textbf{NODE1}          & \textbf{PWC}      & 6.04                      & 7.61                      & 8.26                      & 28.97  & 60.21                     & 95.12   & 0.11   & 1.64                      & 7.15   \\
                            & \textbf{PWA}      & 3.12                      & 4.24                      & 4.89                      & 12.67  & 13.60                     & 14.49   & 0.03   & 0.04                      & 0.04   \\
                            & \textbf{Sig.}      & 18.38                     & 26.35                     & 33.16                     & 20.82  & 23.33                     & 26.79   & 72.75  & 76.83                     & 82.47  \\
    \bottomrule
\end{tabular}

\label{tab:times}

\end{table}
\begin{figure}[htb]
    \centering
        \begin{subfigure}[Locally non-Lipschitz, nonlinear concrete dynamical model.]{
        \centering
        \includegraphics[width=0.22\textwidth]{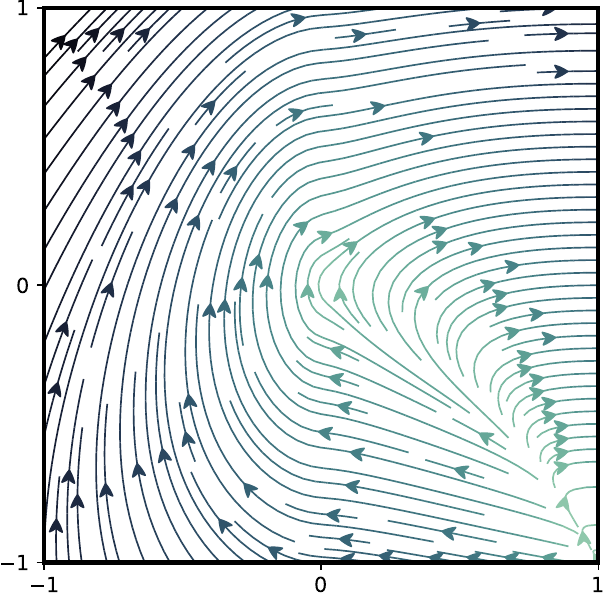}
        \label{fig:concrete}}
    \end{subfigure}
    \begin{subfigure}[Neural PWC abstraction with associated polyhedral partitioning.]{  
        \centering 
        \includegraphics[width=0.22\textwidth]{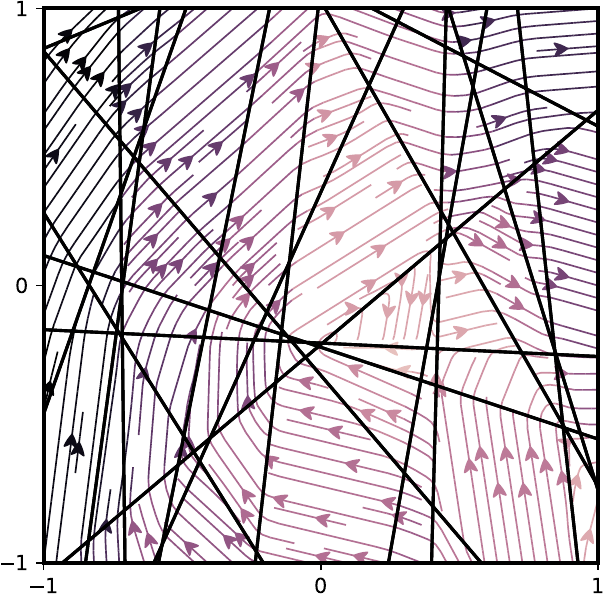}
        \label{fig:nl2-pwc}}
    \end{subfigure}
    \begin{subfigure}[Neural PWA abstraction with associated polyhedral partitioning.]{  
        \centering 
        \includegraphics[width=0.22\textwidth]{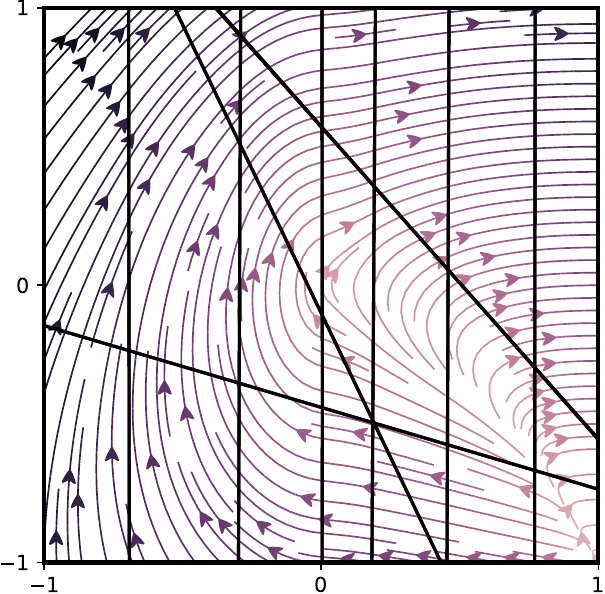}
        \label{fig:nl2-pwa}}
    \end{subfigure}
    \begin{subfigure}[Neural nonlinear (sigmoidal) abstraction.]{  
        \centering 
        \includegraphics[width=0.22\textwidth]{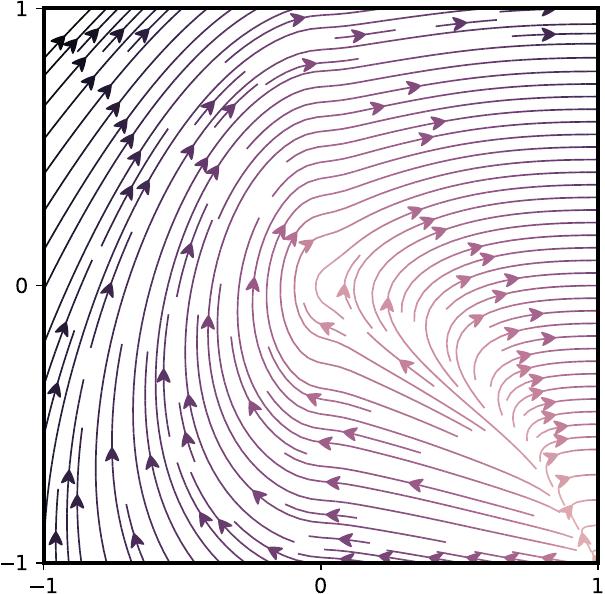}
        \label{fig:nl2-nl}}
    \end{subfigure}
    \caption{Visualisation of Neural Abstractions (underlying flow and relevant partitioning) from piecewise constant (PWC, b), piecewise affine (PWA, c) and sigmoidal (d) templates of a concrete model (a) that does not exhibit local Lipschitz continuity (Non-Lipschitz 2 model). }
    \label{fig:nl2-abs}
\end{figure}
\begin{figure}[htb]
    \centering
    \begin{subfigure}[Flowpipe propagation of neural PWC model.]{  
        \centering 
        \includegraphics[width=0.31\textwidth]{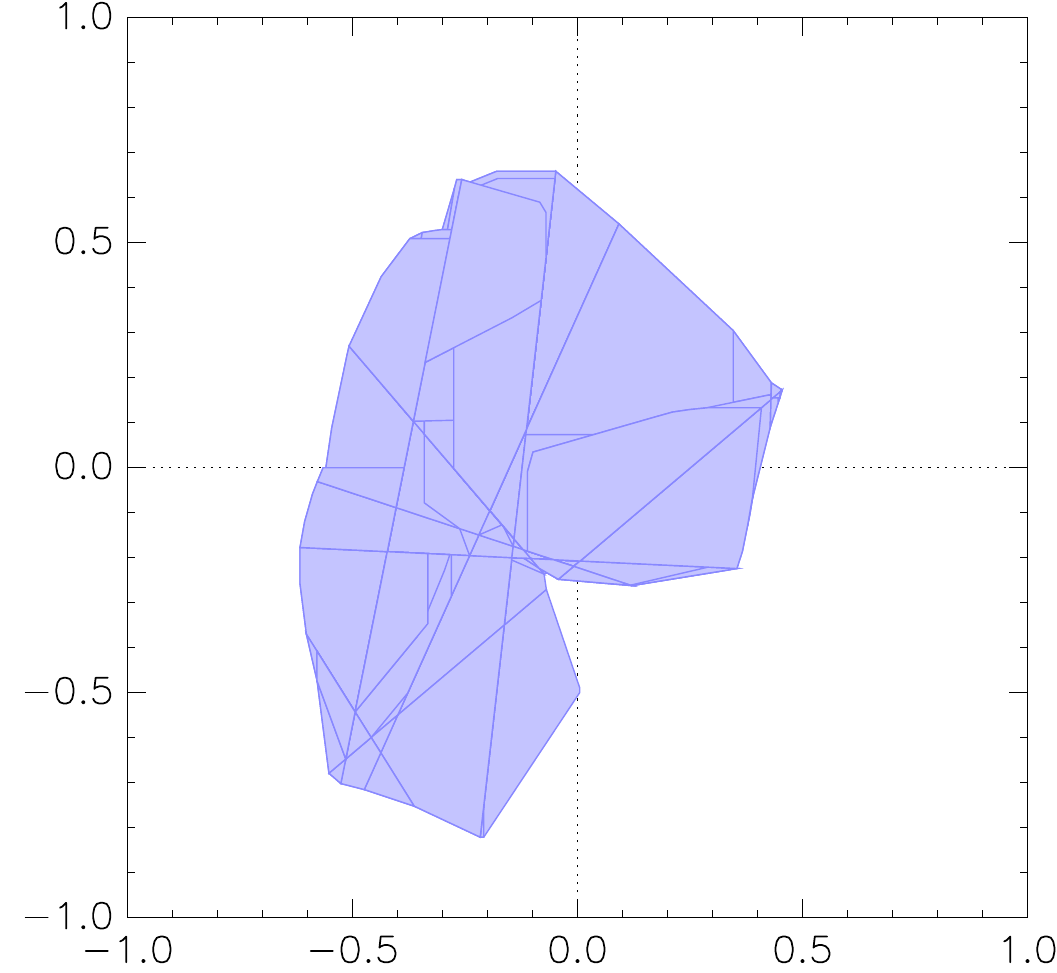}
        \label{fig:nl2-pwc-flow}}
    \end{subfigure}
    \begin{subfigure}[Flowpipe propagation of neural PWA model.]{  
        \centering 
        \includegraphics[width=0.31\textwidth]{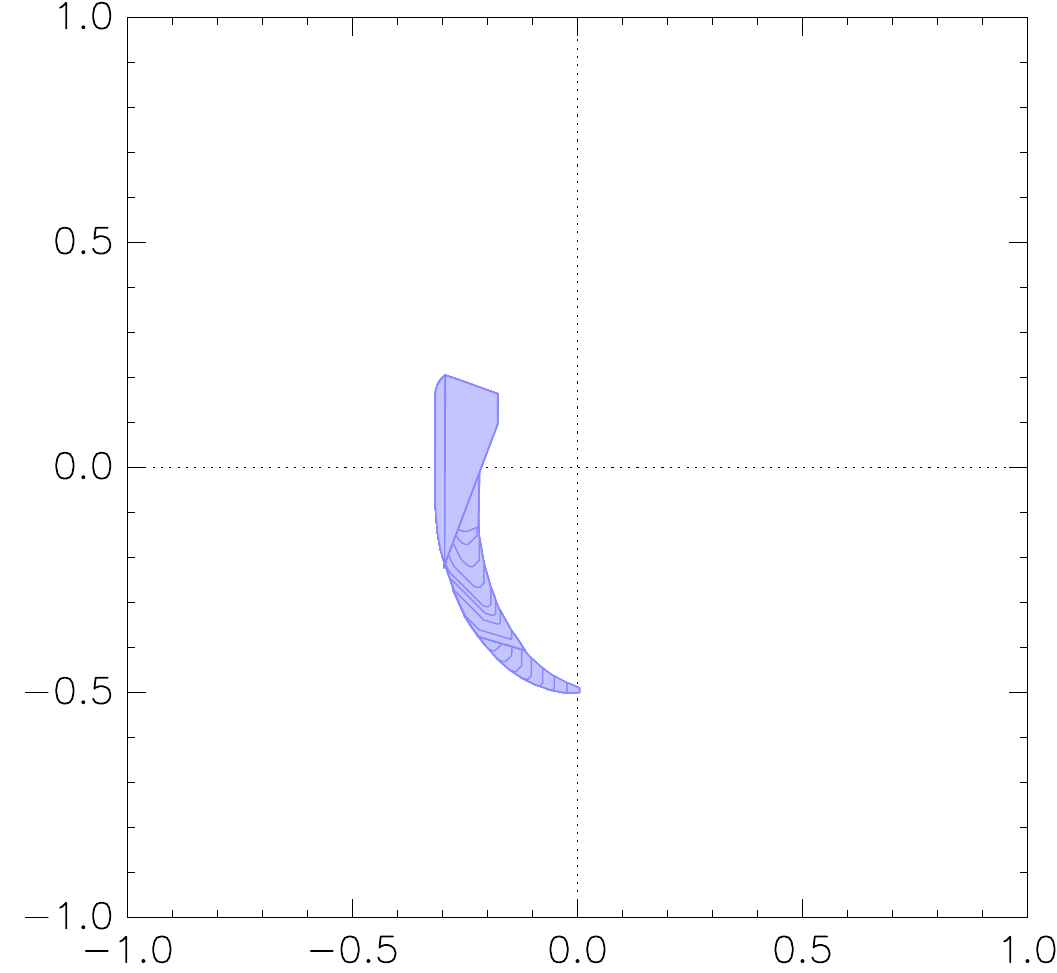}
        \label{fig:nl2-pwa-flow}}
    \end{subfigure}
    \begin{subfigure}[Flowpipe propagation of neural sigmoidal model.]{  
        \centering 
        \includegraphics[width=0.31\textwidth]{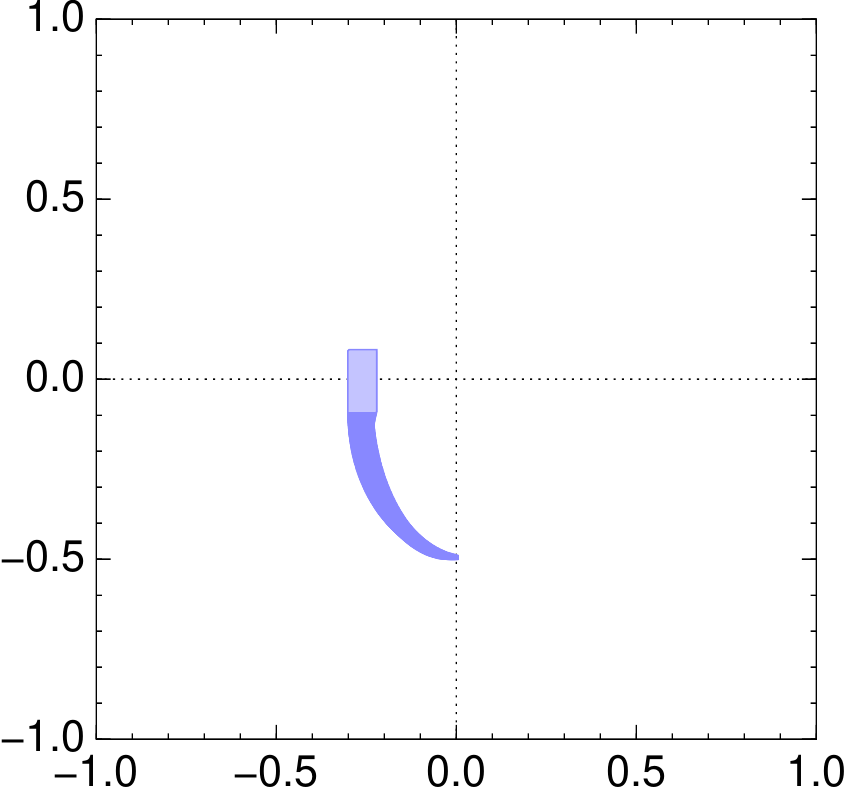}
        \label{fig:nl2-nl-flow}}
    \end{subfigure}
    \caption{Flowpipe (i.e., reach sets) propagation for the Non-Lipschitz 2 model using neural abstraction templates and corresponding verification tools. Flowpipe propagation is performed over the same initial set over the same time horizon (1 second).}
    \label{fig:nl2-flow}
\end{figure}

The first set of benchmarks we consider \emph{do not} exhibit local Lipschitz continuity, meaning they violate the working assumptions of state of the art safety verification tools such as Flow* and GoTube \cite{gruenbacher2022}, and are in general challenging to analyse computationally. Three of the benchmarks are taken from \cite{abate2022neural}, and are shown in the first three rows of \cref{tab:results-tables}. We introduce two new higher dimensional benchmarks, \emph{Water Tank 4D} and \emph{Water Tank 6D}, to demonstrate the ability of our approach to scale to higher-dimensional models with up to six continuous variables. These two additional models are also not locally-Lipschitz continuous. The equations of the dynamics of these models are in~\cref{sec:app-a}. 

For each benchmark, we synthesise an abstraction using one of the proposed templates: piecewise constant, piecewise affine and sigmoidal, and perform flowpipe (i.e., reach sets) propagation over a 1-second time horizon using the appropriate tool\footnote{For the \emph{Water Tank 6D}, we only use the piecewise affine template, as this abstraction performs best in the similar but smaller \emph{Water Tank 4D} experiments.}. 
We leverage our procedure's dependence on random seeding and its low computational cost: for each experiment we initialise multiple runs (namely, four) in parallel. When the first of these returns successfully, we discard the remaining and record the result. For statistical and reproducibility reasons, we repeat each experiment 10 times, and present the mean ($\mu$), max and min over these runs.

We present the salient features of the abstractions in \cref{tab:results-tables}. 
In particular, we present the architecture of the neural network (the number of neurons in the hidden layers); the size of the resulting abstraction in terms of the number of modes $M$ (nonlinear abstractions consist of a single mode) and the 1-norm of the error bound $\epsilon$ -- a vector representing the error bound in each dimension -- for each abstraction. We note that the $\epsilon$ presented is the \emph{global} upper bound to the abstraction error, and does not account for any error refinement achieved by the procedure detailed in \cref{sec:error-refine}. In practice the mode-wise error is often much lower than the global upper bound (see~\cref{sec:app-b} for results on this), which enables us to study higher dimensional models and employ the less expressive piecewise constant templates successfully. The notable exception to this is for \emph{Water Tank 4D}, for which piecewise constant abstractions regularly consist of a very small number of modes, despite a relatively high error: this indicates that the gradient-free learning procedure performs less well for this higher dimensional model, and suggests that neural-based piecewise constant templates are more suitable to smaller dimensional models.

Piecewise constant abstractions perform least well in terms of achieved error bound and in general require larger networks (cf. column $W$ in \cref{tab:results-tables}) and more modes (col. $M$) in the resulting hybrid automaton: this is unsurprising, particularly when compared to the ``higher-order'' piecewise affine abstractions. 
Meanwhile, piecewise affine and simgoidal templates perform more comparably to each other, though sigmoidal templates do achieve slightly better error bounds while using significantly smaller neural networks.

We also present a breakdown on the time spent within each stage of synthesis --training ($T_L$) and certification ($T_C$)) -- in \cref{tab:times}. We prioritised learning speed for the piecewise constant abstractions rather than comprehensive learning, seeking to ensure learning times that were comparable to the other two templates. This is also because we expect safety verification for piecewise constant dynamics to be quite fast. This means that better errors might be achievable for piecewise constant abstractions, however with significantly greater learning times.

We do not perform explicit safety verification using abstract models here, as selecting regions of bad states is arbitrary given that all the abstractions depend on an error bound. Instead, as anticipated earlier, we do perform flowpipe (i.e., reach sets) propagation for each abstraction over a time horizon of 1 second, and present the obtained computation time ($T_f$) in \cref{tab:times}. This column highlights the aforementioned trade-off between abstraction templates: flowpipes can be calculated for piecewise constant abstractions (in general) significantly faster than for the other two in lower dimensions. However, for higher-dimensional or more complex models, the flowpipe computation is much slower: this is due to the abstraction error ($\epsilon$) being greater, making flowpipe computation more difficult. Meanwhile, despite the improved accuracy, the sigmoidal abstractions require more computation for flowpipe propagation due to the increase in model complexity (non-linearity). 

The flowpipe propagation is run with a 500 second timeout, and may not terminate successfully - e.g., Flow* is sometimes unable to compute the whole flowpipe. Across all experiments (including repetitions), the flowpipe propagation is unsuccessful \emph{only three times}: twice for the sigmoidal templates for \emph{Non-Lipschitz 2}, and once for the piecewise constant templates for \emph{WaterTank 4D}. We also set an overall timeout of 1800 seconds on the whole procedure (including abstraction synthesis and flowpipe propagation). \emph{Only four} experiments failed to complete before this timeout: once for the piecewise constant template for \emph{Water Tank 4D}, and three times for the \emph{Water Tank 6D} experiments. We emphasise that these outcomes highlight the robustness and `practical completeness' of the end-to-end procedure. The numerical outcomes shown in the tables exclude those few experiments that time out.

The abstractions for the \emph{Non-Lipschitz 2} model are illustrated in Fig. \ref{fig:nl2-abs}, with the corresponding polyhedral partitioning (for piecewise constant and -affine abstractions) and obtained (locally) approximated vector fields.  
The results of the corresponding flowpipe (i.e., reach sets) propagation are then depicted in Fig. \ref{fig:nl2-flow}. This figure illustrates further the sorts of safety verification tasks that can be completed using each abstraction type: challenging verification tasks, requiring enhanced precision ($\epsilon$), should be attempted using piecewise affine or nonlinear templates, whereas simpler tasks might be formally verified more efficiently through safe piecewise constant neural templates.

\subsection{Abstraction of Neural ODEs}
The results presented in \cref{tab:results-tables} include a model \emph{(NODE1)} that is Lipschitz continuous, but whose dynamics are described by a neural network. These kinds of models are known as neural ODEs (NODEs), and have become a widely studied tool across machine learning and verification \cite{nips/ChenRBD18}.  Here, we discuss neural ODEs as an additional use-case for neural abstractions.

Neural ODEs are commonly trained to approximate real physical models based on sample data. However, it is likely that these models are over-parameterised, making their reachability analysis difficult. For the experiment, we train a neural ODE consisting of a feed-forward network with three hidden layers of hyperbolic tangent activations,  based on data from an underlying two-dimensional model. Then, as with the previous experiments, we construct abstractions of this neural ODE with three templates and perform flowpipe propagation. The longest computation time is for the sigmoidal template at 138 s; in contrast, we provide the concrete neural ODE to Flow* with the same settings, which takes 205 s. It is clear that neural abstractions can be beneficial in improving computational efficiency for flowpipe propagation of neural ODEs at the cost of the error bound $\epsilon$. We consider this to be an alternative approach to reducing the Taylor model with Flow*, but instead results in a model which is permanently easier to analyse.

Finally, we note that we do not use state-of-the-art tools for neural ODE reachability analysis here, such as GoTube \cite{gruenbacher2022} or NNV \cite{manzanaslopez2022ReachabilityAnalysisGeneral}. These tools cannot perform reachability analysis on neural abstractions, as they do not account for the non-determinism introduced by the abstraction error. Thus, in order to make a fair comparison we use Flow*, since to the best of our knowledge, no tool specialised to neural ODEs can also handle nonlinear neural abstractions directly. Our results are promising in light of research interest in neural ODEs, and warrant further interest in the reachability analysis of neural abstractions.

\section{Conclusions}
This work builds on recent literature that employs neural networks as piecewise affine formal abstractions of dynamical models. We extend these abstractions with new neural-based templates which can be cast as models with different, but well studied, semantics -- piecewise constant and nonlinear (sigmoidal). After presenting a workflow for the construction of these abstractions, including templating, training and certification, we study the abstractions via equivalent models that are analysable by existing verification technologies. Using existing tools that implement these technologies, we show the advantages of abstractions of different semantics with regards to their precision and ease to analyse, which indicates their suitability for different verification tasks. 

We improve on existing procedures for synthesising neural abstractions, allowing for abstraction of higher dimensional models. In addition, we demonstrate that neural abstractions of any template can be used to abstract complex neural ODEs, enabling more efficient, though coarser, reachability-based analysis. These results are promising in light of growing interest in neural ODEs; future work in this area should consider extending tools specialised in reachability of neural ODEs to neural abstractions by incorporating the appropriate non-determinism.

\paragraph{Acknowledgments} Alec is grateful for the support of the EPSRC
Centre for Doctoral Training in Autonomous Intelligent Machines and Systems (EP/S024050/1).

\clearpage
    
\appendix
\section{Benchmarks of NonLinear Dynamical Systems}
\label{sec:app-a}
For each dynamical model, we report the vector field $f: \mathbb{R}^n \to \mathbb{R}^n$ and the spatial domain $\mathcal{X}$ over which the abstraction is performed. 

\bigskip 

\textbf{Water Tank}
\begin{equation}
\begin{cases}
  \dot{x} = 1.5 - \sqrt{x} \\
    \mathcal{X}_0 = [0, 0.01]  \\
    \mathcal{X} = [0, 2]  
\end{cases}
\end{equation}

\textbf{Non-Lipschitz Vector Field 1 (NL1)}
\begin{equation}
\begin{cases}
 \dot{x} = y  \\
 \dot{y} = \sqrt{x} \\
 \mathcal{X}_0 = [0, 0.01] \times [0, 0.01],\\
 \mathcal{X} = [0, 1] \times [-1, 1],\\
 
\end{cases}
\end{equation}

\textbf{Non-Lipschitz Vector Field 2 (NL2)}
\begin{equation}
\begin{cases}
 \dot{x} = x^2 + y  \\
 \dot{y} = \sqrt[3]{x^2} -x, \\
 \mathcal{X} = [-1, 1]^2, \\
 \mathcal{X}_0 =  [-0.005, 0.005] \times [-0.5, -0.49] \\
\end{cases}
\end{equation}

\textbf{Water Tank 4D}
\begin{equation}
\begin{cases}
  \dot{x_0} = 0.2 - \sqrt{x_0} \\
  \dot{x_1} =  -x_1 \\
  \dot{x_2} =  -x_2 \\
  \dot{x_3} =  -0.25(x_0 + x_1 + x_2 + x_3) \\
  \mathcal{X} = [0, 1]^4 \\
    \mathcal{X}_0 = [0, 0.01] \times [0.8, 0.81] \times [0.8, 0.81] \times [0.8, 0.81] 
\end{cases}
\end{equation}

\textbf{Water Tank 6D}
\begin{equation}
\begin{cases}
   \dot{x_0} = 0.2 - \sqrt{x_0} \\
   \dot{x_1} =  -x_1 \\
   \dot{x_2} =  -x_2 \\
   \dot{x_3} =  -x_3 \\
   \dot{x_4} =  -x_4 \\
   \dot{x_5} =  -\frac{1}{6}(x_0 + x_1 + x_2 + x_3 + x_4 + x_5) \\
   \mathcal{X} = [0, 1]^6 \\
    \mathcal{X}_0 = [0, 0.01] \times [0.8, 0.81] \times [0.8, 0.81]  \times \\ [0.8, 0.81] \times[0.7, 0.71] \times[0.65, 0.66]   
\end{cases}
\end{equation}

\section{Experiments on Error Refinement Scheme}
\label{sec:app-b}

We consider the effect of the error refinement scheme proposed in \cref{sec:error-refine}, over a single benchmark (\emph{Non-Lipschitz 2}) for 10 repeated runs and no parallelism (i.e., each run consists of only a single attempt
) with and without the error-refinement scheme. We consider first a piecewise affine template, using a network architecture of two hidden layers with 14 and 12 neurons respectively, and a piecewise constant template with a single hidden layer of 20 neurons. These results are shown in \cref{tab:error-tab}, where we report the \textit{mean $(\mu)$} certification time $\bar{T}_C$ and flowpipe propagation time $\bar{T}_f$. We consider any time spent in SMT-solving to be certification time. The time spent in the learner, $T_L$ is not reported as this is not impacted by the refinement scheme.

It is clear that for both templates, the error refinement provides a significant decrease in flowpipe propagation time, at the cost of a minor increase in certification time. In addition to this, it significantly increases the usability of piecewise affine abstractions for SpaceEx with twice as many successfully terminating. We note that runs are considered unsuccessful if they do not terminate successfully from SpaceEx, or within a 500 second timeout, as in previous experiments. Finally, we present the average error bound over all modes in an abstraction, averaged over all successful runs. 

From the reported outcomes, it is clear that the error refinement is able to significantly refine the error for modes in the abstraction, resulting in a more precise abstractions and thus more accurate flowpipe propagation. This error refinement technique is promising, as it can enable the approach to scale to higher-dimensional and more complex models more easily. 

\begin{table}[htb]
    \centering
    \caption{The impact of error refinement on certification time $T_C$, flowpipe propagation time $T_f$, and mean abstraction error over modes $||\epsilon||_1$, as well as the effect on success rate (out of 10 reapeats) of flowpipe propagation for the Non-Lipschitz 2 benchmark for piecewise affine (pwa) and piecewise constant models. The results shown for $T_C$, $T_f$ and $||\epsilon||_1$ are averaged over all successful runs, which we denote using bar notation (rather than $\mu$ as above).}
\begin{tabular}{@{} llrrrr @{}}
    \toprule
    \textbf{Template} & \textbf{Error Refinement}& $\bar{T}_C$ & $\bar{T}_f$ & $ ||\bar{\epsilon}||_1$   & Success Rate \\
    \midrule
    \textbf{PWC}      & \textbf{Without}            & 0.92        & 16.45       & 1.89                      & 1.00         \\
                      & \textbf{With }            & 1.27        & 7.87        & 1.20                      & 1.00         \\
    
    \textbf{PWA}      & \textbf{Without}            & 14.54       & 54.17       & 0.14                      & 0.30         \\
                      & \textbf{With }            & 16.46       & 39.67       & 0.07                      & 0.60         \\
    \bottomrule
\end{tabular}

    \label{tab:error-tab}
\end{table}

\clearpage
\bibliographystyle{splncs04}
\bibliography{main}

\begin{thebibliography}{10}
\providecommand{\url}[1]{\texttt{#1}}
\providecommand{\urlprefix}{URL }
\providecommand{\doi}[1]{https://doi.org/#1}

\bibitem{abate2022neural}
Abate, A., Edwards, A., Giacobbe, M.: Neural abstractions. In: Thirty-Sixth
  Conference on Neural Information Processing Systems (2022)

\bibitem{hybrid/Althoff13}
Althoff, M.: Reachability analysis of nonlinear systems using conservative
  polynomialization and non-convex sets. In: {HSCC}. pp. 173--182. {ACM} (2013)

\bibitem{cdc/AlthoffSB08}
Althoff, M., Stursberg, O., Buss, M.: Reachability analysis of nonlinear
  systems with uncertain parameters using conservative linearization. In:
  {CDC}. pp. 4042--4048. {IEEE} (2008)

\bibitem{alur1995AlgorithmicAnalysisHybrid}
Alur, R., Courcoubetis, C., Halbwachs, N., Henzinger, T., Ho, P.H., Nicollin,
  X., Olivero, A., Sifakis, J., Yovine, S.: The algorithmic analysis of hybrid
  systems. Theoretical Computer Science  \textbf{138}(1),  3--34 (1995)

\bibitem{alur1996AutomaticSymbolicVerification}
Alur, R., Henzinger, T., Ho, P.H.: Automatic symbolic verification of embedded
  systems. IEEE Transactions on Software Engineering  \textbf{22}(3),  181--201
  (1996)

\bibitem{871304}
Alur, R., Henzinger, T., Lafferriere, G., Pappas, G.: Discrete abstractions of
  hybrid systems. Proceedings of the IEEE  \textbf{88}(7),  971--984 (2000)

\bibitem{hybrid/AsarinD04}
Asarin, E., Dang, T.: Abstraction by projection and application to multi-affine
  systems. In: {HSCC}. Lecture Notes in Computer Science, vol.~2993, pp.
  32--47. Springer (2004)

\bibitem{hybrid/AsarinDG03}
Asarin, E., Dang, T., Girard, A.: Reachability analysis of nonlinear systems
  using conservative approximation. In: {HSCC}. Lecture Notes in Computer
  Science, vol.~2623, pp. 20--35. Springer (2003)

\bibitem{acta/AsarinDG07}
Asarin, E., Dang, T., Girard, A.: Hybridization methods for the analysis of
  nonlinear systems. Acta Informatica  \textbf{43}(7),  451--476 (2007)

\bibitem{ijcai/BacciG021}
Bacci, E., Giacobbe, M., Parker, D.: Verifying reinforcement learning up to
  infinity. In: {IJCAI}. pp. 2154--2160. ijcai.org (2021)

\bibitem{Bak_Bogomolov_Duggirala_Gerlach_Potomkin_2021}
Bak, S., Bogomolov, S., Duggirala, P.S., Gerlach, A.R., Potomkin, K.:
  Reachability of black-box nonlinear systems after koopman operator
  linearization. IFAC-PapersOnLine  \textbf{54}(5),  253--258 (2021), 7th IFAC
  Conference on Analysis and Design of Hybrid Systems ADHS 2021

\bibitem{hybrid/BakBHJP16}
Bak, S., Bogomolov, S., Henzinger, T.A., Johnson, T.T., Prakash, P.: Scalable
  static hybridization methods for analysis of nonlinear systems. In: {HSCC}.
  pp. 155--164. {ACM} (2016)

\bibitem{bogomolov2017CounterexampleGuidedRefinementTemplate}
Bogomolov, S., Frehse, G., Giacobbe, M., Henzinger, T.A.:
  Counterexample-{{Guided Refinement}} of {{Template Polyhedra}}. In: Legay,
  A., Margaria, T. (eds.) Tools and {{Algorithms}} for the {{Construction}} and
  {{Analysis}} of {{Systems}}, vol. 10205, pp. 589--606. {Springer Berlin
  Heidelberg}, {Berlin, Heidelberg} (2017)

\bibitem{formats/BogomolovGHK17}
Bogomolov, S., Giacobbe, M., Henzinger, T.A., Kong, H.: Conic abstractions for
  hybrid systems. In: {FORMATS}. Lecture Notes in Computer Science, vol. 10419,
  pp. 116--132. Springer (2017)

\bibitem{nips/ChenRBD18}
Chen, T.Q., Rubanova, Y., Bettencourt, J., Duvenaud, D.: Neural ordinary
  differential equations. In: NeurIPS. pp. 6572--6583 (2018)

\bibitem{rtss/ChenAS12}
Chen, X., {\'{A}}brah{\'{a}}m, E., Sankaranarayanan, S.: Taylor model flowpipe
  construction for non-linear hybrid systems. In: {RTSS}. pp. 183--192. {IEEE}
  Computer Society (2012)

\bibitem{cav/ChenAS13}
Chen, X., {\'{A}}brah{\'{a}}m, E., Sankaranarayanan, S.: Flow*: An analyzer for
  non-linear hybrid systems. In: {CAV}. Lecture Notes in Computer Science,
  vol.~8044, pp. 258--263. Springer (2013)

\bibitem{tecs/ChenMS17}
Chen, X., Mover, S., Sankaranarayanan, S.: Compositional relational abstraction
  for nonlinear hybrid systems. {ACM} Trans. Embed. Comput. Syst.
  \textbf{16}(5s),  187:1--187:19 (2017)

\bibitem{rtss/0002S16}
Chen, X., Sankaranarayanan, S.: Decomposed reachability analysis for nonlinear
  systems. In: {RTSS}. pp. 13--24. {IEEE} Computer Society (2016)

\bibitem{clarke2000CounterexampleguidedAbstractionRefinement}
Clarke, E., Grumberg, O., Jha, S., Lu, Y., Veith, H.: Counterexample-guided
  abstraction refinement. In: Emerson, E.A., Sistla, A.P. (eds.) Computer Aided
  Verification. pp. 154--169. {Springer Berlin Heidelberg}, {Berlin,
  Heidelberg} (2000)

\bibitem{hybrid/DangMT10}
Dang, T., Maler, O., Testylier, R.: Accurate hybridization of nonlinear
  systems. In: {HSCC}. pp. 11--20. {ACM} (2010)

\bibitem{hybrid/DangT11}
Dang, T., Testylier, R.: Hybridization domain construction using curvature
  estimation. In: {HSCC}. pp. 123--132. {ACM} (2011)

\bibitem{demoura2008Z3EfficientSMT}
{de Moura}, L., Bj{\o}rner, N.: Z3: {{An Efficient SMT Solver}}. In: Hutchison,
  D., Kanade, T., Kittler, J., Kleinberg, J.M., Mattern, F., Mitchell, J.C.,
  Naor, M., Nierstrasz, O., Pandu~Rangan, C., Steffen, B., Sudan, M.,
  Terzopoulos, D., Tygar, D., Vardi, M.Y., Weikum, G., Ramakrishnan, C.R.,
  Rehof, J. (eds.) Tools and {{Algorithms}} for the {{Construction}} and
  {{Analysis}} of {{Systems}}, vol.~4963, pp. 337--340. {Springer Berlin
  Heidelberg}, {Berlin, Heidelberg} (2008)

\bibitem{hybrid/DuttaCS19}
Dutta, S., Chen, X., Sankaranarayanan, S.: Reachability analysis for neural
  feedback systems using regressive polynomial rule inference. In: {HSCC}. pp.
  157--168. {ACM} (2019)

\bibitem{cav/FanQM0D16}
Fan, C., Qi, B., Mitra, S., Viswanathan, M., Duggirala, P.S.: Automatic
  reachability analysis for nonlinear hybrid models with {C2E2}. In: {CAV}
  {(1)}. Lecture Notes in Computer Science, vol.~9779, pp. 531--538. Springer
  (2016)

\bibitem{frehse2008PHAVerAlgorithmicVerification}
Frehse, G.: {{PHAVer}}: Algorithmic verification of hybrid systems past
  {{HyTech}}. International Journal on Software Tools for Technology Transfer
  \textbf{10}(3),  263--279 (Jun 2008)

\bibitem{cav/FrehseGDCRLRGDM11}
Frehse, G., Guernic, C.L., Donz{\'{e}}, A., Cotton, S., Ray, R., Lebeltel, O.,
  Ripado, R., Girard, A., Dang, T., Maler, O.: Spaceex: Scalable verification
  of hybrid systems. In: {CAV}. Lecture Notes in Computer Science, vol.~6806,
  pp. 379--395. Springer (2011)

\bibitem{frehse2013FlowpipeApproximationClustering}
Frehse, G., Kateja, R., Le~Guernic, C.: Flowpipe approximation and clustering
  in space-time. In: Proceedings of the 16th International Conference on
  {{Hybrid}} Systems: Computation and Control - {{HSCC}} '13. p.~203. {ACM
  Press}, {Philadelphia, Pennsylvania, USA} (2013)

\bibitem{gao2013DRealSMTSolver}
Gao, S., Kong, S., Clarke, E.M.: {{dReal}}: {{An SMT Solver}} for {{Nonlinear
  Theories}} over the {{Reals}}. In: Hutchison, D., Kanade, T., Kittler, J.,
  Kleinberg, J.M., Mattern, F., Mitchell, J.C., Naor, M., Nierstrasz, O.,
  Pandu~Rangan, C., Steffen, B., Sudan, M., Terzopoulos, D., Tygar, D., Vardi,
  M.Y., Weikum, G., Bonacina, M.P. (eds.) Automated {{Deduction}} \textendash{}
  {{CADE-24}}, vol.~7898, pp. 208--214. {Springer Berlin Heidelberg}, {Berlin,
  Heidelberg} (2013)

\bibitem{aaai/GruenbacherHLCS21}
Gruenbacher, S., Hasani, R.M., Lechner, M., Cyranka, J., Smolka, S.A., Grosu,
  R.: On the verification of neural odes with stochastic guarantees. In:
  {AAAI}. pp. 11525--11535. {AAAI} Press (2021)

\bibitem{gruenbacher2022}
Gruenbacher, S., Lechner, M., Hasani, R., Rus, D., Henzinger, T.A., Smolka, S.,
  Grosu, R.: Gotube: Scalable stochastic verification of continuous-depth
  models. In: {AAAI} (2022)

\bibitem{DBLP:conf/lics/Henzinger96}
Henzinger, T.A.: The theory of hybrid automata. In: {LICS}. pp. 278--292.
  {IEEE} Computer Society (1996)

\bibitem{henzinger1997HYTECHModelChecker}
Henzinger, T.A., Ho, P.H., {Wong-Toi}, H.: {{HYTECH}}: A model checker for
  hybrid systems. International Journal on Software Tools for Technology
  Transfer  \textbf{1}(1-2),  110--122 (Dec 1997)

\bibitem{hybrid/HenzingerW95}
Henzinger, T.A., Wong{-}Toi, H.: Linear phase-portrait approximations for
  nonlinear hybrid systems. In: Hybrid Systems. Lecture Notes in Computer
  Science, vol.~1066, pp. 377--388. Springer (1995)

\bibitem{tecs/HuangFLC019}
Huang, C., Fan, J., Li, W., Chen, X., Zhu, Q.: Reachnn: Reachability analysis
  of neural-network controlled systems. {ACM} Trans. Embed. Comput. Syst.
  \textbf{18}(5s),  106:1--106:22 (2019)

\bibitem{cav/IvanovCWAPL21}
Ivanov, R., Carpenter, T.J., Weimer, J., Alur, R., Pappas, G.J., Lee, I.:
  Verisig 2.0: Verification of neural network controllers using taylor model
  preconditioning. In: {CAV} {(1)}. Lecture Notes in Computer Science, vol.
  12759, pp. 249--262. Springer (2021)

\bibitem{cdc/KekatosFF17}
Kekatos, N., Forets, M., Frehse, G.: Constructing verification models of
  nonlinear simulink systems via syntactic hybridization. In: {CDC}. pp.
  1788--1795. {IEEE} (2017)

\bibitem{khalil2002NonlinearSystems}
Khalil, H.K.: Nonlinear Systems. {Prentice Hall}, {Upper Saddle River, N.J},
  3rd ed edn. (2002)

\bibitem{hsb/KongBBGHJS16}
Kong, H., Bartocci, E., Bogomolov, S., Grosu, R., Henzinger, T.A., Jiang, Y.,
  Schilling, C.: Discrete abstraction of multiaffine systems. In: {HSB}.
  Lecture Notes in Computer Science, vol.~9957, pp. 128--144 (2016)

\bibitem{tacas/KongGCC15}
Kong, S., Gao, S., Chen, W., Clarke, E.M.: dreach: {\(\delta\)}-reachability
  analysis for hybrid systems. In: {TACAS}. Lecture Notes in Computer Science,
  vol.~9035, pp. 200--205. Springer (2015)

\bibitem{formats/LiBB20}
Li, D., Bak, S., Bogomolov, S.: Reachability analysis of nonlinear systems
  using hybridization and dynamics scaling. In: {FORMATS}. Lecture Notes in
  Computer Science, vol. 12288, pp. 265--282. Springer (2020)

\bibitem{mackay2003InformationTheoryInference}
MacKay, D.J.C.: Information Theory, Inference, and Learning Algorithms.
  {Cambridge University Press}, {Cambridge, UK ; New York} (2003)

\bibitem{cav/MajumdarZ12}
Majumdar, R., Zamani, M.: Approximately bisimilar symbolic models for digital
  control systems. In: {CAV}. Lecture Notes in Computer Science, vol.~7358, pp.
  362--377. Springer (2012)

\bibitem{manzanaslopez2022ReachabilityAnalysisGeneral}
Manzanas~Lopez, D., Musau, P., Hamilton, N.P., Johnson, T.T.: Reachability
  analysis of a general class of neural ordinary differential equations. In:
  Bogomolov, S., Parker, D. (eds.) Formal Modeling and Analysis of Timed
  Systems. pp. 258--277. {Springer International Publishing}, {Cham} (2022)

\bibitem{pyswarmsJOSS2018}
Miranda, L.J.V.: {P}y{S}warms, a research-toolkit for {P}article {S}warm
  {O}ptimization in {P}ython. Journal of Open Source Software  \textbf{3}
  (2018)

\bibitem{cav/MoverCGIT21}
Mover, S., Cimatti, A., Griggio, A., Irfan, A., Tonetta, S.: Implicit
  semi-algebraic abstraction for polynomial dynamical systems. In: {CAV} {(1)}.
  Lecture Notes in Computer Science, vol. 12759, pp. 529--551. Springer (2021)

\bibitem{pytorch}
Paszke, A., Gross, S., Massa, F., Lerer, A., Bradbury, J., Chanan, G., Killeen,
  T., Lin, Z., Gimelshein, N., Antiga, L., Desmaison, A., K{\"{o}}pf, A., Yang,
  E.Z., DeVito, Z., Raison, M., Tejani, A., Chilamkurthy, S., Steiner, B.,
  Fang, L., Bai, J., Chintala, S.: {PyTorch:} an imperative style,
  high-performance deep learning library. In: NeurIPS. pp. 8024--8035 (2019)

\bibitem{pola2008ApproximatelyBisimilarSymbolic}
Pola, G., Girard, A., Tabuada, P.: Approximately bisimilar symbolic models for
  nonlinear control systems. arXiv:0706.0246 [math]  (Jan 2008)

\bibitem{tac/PrabhakarD015}
Prabhakar, P., Dullerud, G.E., Viswanathan, M.: Stability preserving
  simulations and bisimulations for hybrid systems. {IEEE} Trans. Autom.
  Control.  \textbf{60}(12),  3210--3225 (2015)

\bibitem{cav/PrabhakarS13}
Prabhakar, P., Soto, M.G.: Abstraction based model-checking of stability of
  hybrid systems. In: {CAV}. Lecture Notes in Computer Science, vol.~8044, pp.
  280--295. Springer (2013)

\bibitem{tacas/RoohiP016}
Roohi, N., Prabhakar, P., Viswanathan, M.: Hybridization based {CEGAR} for
  hybrid automata with affine dynamics. In: {TACAS}. Lecture Notes in Computer
  Science, vol.~9636, pp. 752--769. Springer (2016)

\bibitem{hybrid/Sankaranarayanan11}
Sankaranarayanan, S.: Automatic abstraction of non-linear systems using change
  of bases transformations. In: {HSCC}. pp. 143--152. {ACM} (2011)

\bibitem{cav/SankaranarayananT11}
Sankaranarayanan, S., Tiwari, A.: Relational abstractions for continuous and
  hybrid systems. In: {CAV}. Lecture Notes in Computer Science, vol.~6806, pp.
  686--702. Springer (2011)

\bibitem{sastry1999NonlinearSystems}
Sastry, S.: Nonlinear {{Systems}}, Interdisciplinary {{Applied Mathematics}},
  vol.~10. {Springer New York}, {New York, NY} (1999)

\bibitem{van_der_Schaft_Schumacher_2000}
van~der Schaft, A., Schumacher, H.: An introduction to hybrid dynamical
  systems, Lecture Notes in Control and Information Sciences, vol.~251.
  Springer London, London (2000)

\bibitem{schilling2022}
Schilling, C., Forets, M., Guadalupe, S.: Verification of neural-network
  control systems by integrating taylor models and zonotopes. In: {AAAI} (2022)

\bibitem{solar-lezama2006CombinatorialSketchingFinite}
{Solar-Lezama}, A., Tancau, L., Bodik, R., Seshia, S., Saraswat, V.:
  Combinatorial sketching for finite programs. SIGOPS Oper. Syst. Rev.
  \textbf{40}(5),  404--415 (Oct 2006)

\bibitem{rtss/SotoP20}
Soto, M.G., Prabhakar, P.: Hybridization for stability verification of
  nonlinear switched systems. In: {RTSS}. pp. 244--256. {IEEE} (2020)

\bibitem{tecs/TranCLMJK19}
Tran, H., Cai, F., Lopez, D.M., Musau, P., Johnson, T.T., Koutsoukos, X.D.:
  Safety verification of cyber-physical systems with reinforcement learning
  control. {ACM} Trans. Embed. Comput. Syst.  \textbf{18}(5s),  105:1--105:22
  (2019)

\bibitem{cav/TranYLMNXBJ20}
Tran, H., Yang, X., Lopez, D.M., Musau, P., Nguyen, L.V., Xiang, W., Bak, S.,
  Johnson, T.T.: {NNV:} the neural network verification tool for deep neural
  networks and learning-enabled cyber-physical systems. In: {CAV} {(1)}.
  Lecture Notes in Computer Science, vol. 12224, pp. 3--17. Springer (2020)

\bibitem{amcc/XiangTRJ18}
Xiang, W., Tran, H., Rosenfeld, J.A., Johnson, T.T.: Reachable set estimation
  and safety verification for piecewise linear systems with neural network
  controllers. In: {ACC}. pp. 1574--1579. {IEEE} (2018)

\end{thebibliography}

\end{document}